# Disruption of Saturn's Quasi-Periodic Equatorial Oscillation by the Great Northern Storm

Leigh N. Fletcher[1*], Sandrine Guerlet[2], Glenn S. Orton[3], Richard G. Cosentino[4], Thierry Fouchet[5], Patrick G.J. Irwin[6], Liming Li[7], F. Michael Flasar[4], Nicolas Gorius[4], Raúl Morales-Juberías[8]

1. Department of Physics and Astronomy, University of Leicester, University Road, Leicester, LE1 7RH, UK (leigh.fletcher@leicester.ac.uk)
2. Laboratoire de Meteorologie Dynamique /IPSL, Sorbonne Universités, UPMC Univ Paris 06, CNRS, Paris, France
3. Jet Propulsion Laboratory, California Institute of Technology, 4800 Oak Grove Drive, Pasadena, CA, 91109, USA
4. NASA Goddard Spaceflight Center, Maryland, MD, USA.
5. LESIA, Observatoire de Paris, PSL Research University, CNRS, Sorbonne Universités, UPMC Univ. Paris 06, Univ. Paris Diderot, Sorbonne Paris Cité, 5 place Jules Janssen, 92195 Meudon, France
6. Atmospheric, Oceanic and Planetary Physics, University of Oxford, Parks Road, Oxford, OX1 3PU, UK.
7. University of Houston, Department of Physics, Houston, TX 77004, USA
8. New Mexico Institute of Mining and Technology, Socorro NM 87801, USA

* Corresponding author.

**Observations of planets throughout our Solar System have revealed that the Earth is not alone in possessing natural, inter-annual atmospheric cycles [1-4]. The equatorial middle atmospheres of the Earth, Jupiter and Saturn all exhibit a remarkably similar phenomenon – a vertical, cyclic pattern of alternating temperatures and zonal (east-west) wind regimes that propagate slowly downwards with a well-defined multi-Earth-year period. Earth's Quasi-Biennial Oscillation (QBO, observed in the lower stratospheres with an average period of 28 months) is one of the most regular, repeatable cycles exhibited by our climate system [1], and yet recent work has shown that this regularity can be disrupted by events occurring far away from the equatorial region [5,6], an example of a phenomenon known as atmospheric teleconnection. Here we reveal that Saturn's equatorial Quasi-Periodic Oscillation (QPO, with a ~15-year period) can also be dramatically perturbed. An intense springtime storm erupted at Saturn's northern mid-latitudes in December 2010 [7,8,9], spawning a gigantic hot vortex in the stratosphere at 40°N that persisted for 3 years [10]. Far from the storm, the Cassini temperature measurements showed a dramatic ~10-K cooling in the 0.5-5 mbar range across the entire equatorial region, disrupting the regular QPO pattern and significantly altering the middle-atmospheric wind structure, suggesting an injection of westward momentum into the equatorial wind system from waves generated by the northern storm. Hence, as on Earth, meteorological activity at mid-latitudes can have a profound effect on the regular atmospheric cycles in the tropics, demonstrating that waves can provide horizontal teleconnections between the phenomena shaping the middle atmospheres of giant planets.**

Equatorial oscillations in the stratospheres of Earth, Jupiter, and Saturn manifest vertical patterns of alternating zonal (east-west) axisymmetric wind regimes (zonal jets) and associated temperature anomalies. Earth's Quasi-Biennial Oscillation (QBO) has been observed in wind measurements in the lower stratosphere since the 1950s [1,11,12] and has an average period of 28 months. For the gas giant planets, where direct stratospheric wind measurements are unavailable, the oscillations are inferred from thermal-infrared observations of the 3D temperature structure, which revealed both Jupiter's 4-5-year period Quasi-Quadrennial Oscillation (QQO, [2]) and Saturn's 14.7±0.9-year Quasi-Periodic Oscillation (QPO, [3,4]). By analogy to Earth's tropics, the origin and evolution of these middle-atmospheric cycles may be linked to waves resulting from tropospheric meteorology [1,2,4,13,14]. Convection launches a spectrum of upwardly-propagating atmospheric waves with westward and eastward phase velocities that transport momentum into the stratosphere [15,16]. These waves are damped and preferentially deposit their momentum at levels where the mean zonal flow velocity is near the phase velocity of the wave. Eastward (westward) momentum is absorbed in eastward (westward) shear zones just below the peaks of the zonal winds, accelerating the shear zones downwards to ultimately dissipate near the tropopause. The shear zones filter waves of the same velocity (speed and direction): an eastward shear zone therefore absorbs eastward waves whilst allowing westward waves to propagate into the high stratosphere, and vice versa. Once the shear zone has vanished, newly-generated waves with this velocity can now propagate through to higher altitudes, forming a new shear zone to repeat the cycle [17]. At higher altitude, Earth also displays equatorial semi-annual oscillations in the upper stratosphere and mesosphere that are distinct from the QBO [17]. Disrupting these cycles therefore requires some mechanism to interrupt or modify this wave-induced momentum transport responsible for the stratospheric pattern.

Saturn's QPO and associated equatorial jets were discovered and analysed in observations that lacked good temporal sampling [3,14,18,20-22], but indicated that Saturn's oscillation might be semi-annual with respect to Saturn's year. This was called into question by a comparison of stratospheric thermal structures between Voyager and Cassini data at the same point in Saturn's seasonal cycle (2 QPO cycles apart), which, counter to expectations, showed significant differences in the equatorial temperature contrasts at 1 mbar between the two epochs [23]. This hinted at interannual variability of the QPO. To explore the evolution of this oscillation beyond mere snapshots, we utilise the unprecedented time series provided by the Cassini Composite Infrared Spectrometer (CIRS, 2004-2017, [24]), zonally averaging nadir 15 $cm^{-1}$-resolution spectra spanning 600-1400 $cm^{-1}$ (7.1-16.6 µm) to generate a monthly database between ±30$^o$ latitude (see Methods). Each spectrum was inverted to estimate zonally-averaged stratospheric temperatures (0.5-5 mbar) and hydrocarbon distributions from methane, ethane and acetylene emission features. Limb-viewing observations with improved vertical resolution [21] were used as anchor points for the time-series to confirm consistency with previous studies (see Methods). Interpolation of the time series in Fig. 1 produced animations of temperatures, temperature anomalies and zonal winds (see Methods and Supplemental Information), snapshots of which can be seen in Fig. 2.

Supplemental videos 1-3 and Fig. 3 show the regular downward propagation of the QPO temperature anomalies prior to 2011, both at the equator (averaged over ±5$^o$) and in the extra-tropical regions (±16°) associated with secondary circulation patterns, as on Earth [17]. Seasonal temperature changes due to Saturn's 27° orbital obliquity are evident in comparing 2006 (pre-equinox) to 2015 (post-equinox) in Fig. 2 at latitudes > 10° [20], but the equatorial temperature field is dominated by the QPO. We estimate that the thermal anomalies descend at a rate of 17±1 km/year (0.5±0.1 mm/s) or ~0.3 scale heights/year at 1 mbar, consistent with previous estimates [14,21]. These warm and cool anomalies are separated by approximately a decade of pressure (2-3 pressure scale heights), and the descent speed remains consistent with a semi-annual cycle. The descent is not likely to be rigid, with some distortion of the shape of the QPO as it descends, but we find no measurable distinction between the descent rates of warm and cool anomalies, unlike on Earth where eastward shear zones descend more rapidly than westward shear zones [1].

Cassini arrived at Saturn during northern winter, when the 1-mbar equatorial temperatures in Fig. 1b were at their peak [20], associated with a strong eastward QPO phase [18] and a bright band at Saturn's equator at 7.8 μm [3]. The equator cooled ~20 K at 1 mbar as the shear zone moved downwards, being coolest in 2012-2014 before starting to recover with the onset of a new eastward phase by 2016. The opposite is seen at 5 mbar, where temperatures increased until 2010 but fell from 2010 to 2017 (see Supplemental Information). The removal of the warm equatorial band was evident in ground-based ESO VLT/VISIR and Subaru/COMICS imaging observations at 7.9 and 12.3 μm taken between 2005 and 2016 (Fig. 1b and Supplemental Information), consistent with the switch to the westward QPO phase near 1 mbar (see Methods). However, closer inspection of the 1-mbar temperatures at ±10° latitude in Fig. 1a and 1c showed that they had been stable over time until a dramatic 10-12K cooling event in 2011, which persisted until at least 2014. The 'normal' QPO evolution resumed after 2014, but the sudden 2011-2014 perturbation, symmetric about Saturn's equator in Fig. 2 and restricted to p<5 mbar, was counter to any expectations for the smoothly-oscillating QPO.

In December 2010, a planet-encircling storm had erupted in Saturn's northern mid-latitudes (40°N), with powerful convective plumes lofting ices and gaseous species into the upper troposphere [7-9]. The effects on Saturn's stratosphere provided indirect evidence of vertically-propagating waves transporting energy and momentum into the middle atmosphere [7,10]. Although the tropospheric storm system abated after 6-8 months, the stratospheric aftereffects were evident for more than 3 years, with the formation of an enormous, westward-moving hot vortex (the 'beacon', [10]), with temperatures elevated ~80 K above the quiescent stratosphere at 2 mbar. Both the tropospheric storm, and the stratospheric vortex that it spawned, are likely to have generated substantial wave activity, but their impact on tropical temperatures on both sides of Saturn's equator was unexpected.

Stratospheric winds can be estimated by assuming geostrophic balance and using the latitudinal temperature gradients to integrate zonal winds along cylinders parallel to Saturn's rotation axis. This modified thermal wind equation (rather than integrating in altitude) is required because of the proximity to the equator, where the Coriolis parameter vanishes (see Methods). Figs. 2 and 3 show stratospheric winds relative to the 5-mbar level, and indicate that the strongly-eastward flow observed at the start of Cassini's observing record [18] moved downward from 2004 to 2010. Immediately before the storm onset, a westward phase (relative to the 5-mbar level) was forming at p<0.5 mbar as part of the regular QPO cycle. The 2011 storm eruption initially decelerated the eastward flow throughout the ±10° range near 1-3 mbar, most probably due to an injection of westward momentum, removing the strong eastward windshear near 5 mbar in Fig. 3c. By 2012, when the cooling at ±10° was at its maximum, the windshear had shifted so that weak eastward winds were distributed over the full 0.5-5.0 mbar range, obliterating all signs of the QPO pattern and the weak westward phase that had developed in 2010. However, by 2013 the westward windshear was fully established near 1 mbar (Fig. 3c), such that the westward phase[*] (and the cool equatorial temperatures) prevailed at 1 mbar throughout 2014-2016. This suggests that, despite the 3-year QPO disruption and the abrupt cooling of the ±10° region, the equatorial oscillation was able to re-establish its standard phase progression. The result was that the strong prograde jet observed at the start of the Cassini mission [18] had been considerably weakened as the QPO switched to the opposite phase. Finally, in 2016 a new region of eastward windshear and warmer temperatures can be seen for p<1 mbar in Fig. 3b-c, signalling the onset of the slow switch back to the eastward phase that was seen at the start of Cassini's observations.

An equatorial injection of westward momentum from the northern mid-latitudes is in agreement with both the westward motion of the stratospheric vortex, the anticyclonic nature of the rising storm plumes [25], and with the hypothesis that the beacon formed through interactions of Rossby and gravity waves forced by tropospheric convection (i.e., a source of westward momentum) with the background stratospheric flow [10]. Saturn's middle atmospheric flow is strongly eastward throughout the low latitudes due to the broad equatorial super-rotating jet [26] and the small vertical windshears away from the equator [7, 27], implying an absence of critical surfaces to absorb stationary or westward-moving Rossby waves (i.e., those waves moving west with respect to the mean eastward equatorial flow). This means that any waves radiated from the northern storm in three dimensions were not latitudinally confined and could propagate to the equator where they encountered the eastward shearzone of the QPO pattern in Fig. 3c. The perturbation to the QPO pattern was remarkably symmetric about the equator (Fig. 2), with implications as far south as 20°S – a far wider range of influence for the storm than previously thought. Given that Saturn's equatorial jet is also symmetric, this supports the concept of a wave-mean-flow interaction, with momentum flux deposition maximised in critical regions near ±10° latitude that depend upon the velocity of the background flow. Furthermore, Rossby wave activity is often restricted to low pressures by the static stability [28], favouring the stratospheric vortex as the source of wave-induced westward momentum, rather than the storm itself.

Earth's QBO impacts extra-tropical latitudes via teleconnections, providing a link between weather changes in regions that are geographically separated [29]. Indeed, Earth's QBO was severely disrupted in 2015-16 by an influx of westward momentum by waves emanating from extra-tropical sources in the northern winter hemisphere, forming a new westward layer within the eastward phase [5, 6]. This adjustment was unprecedented in the 66-year QBO record.  The observation of Saturn's QPO disruption during the storm epoch suggests that it possesses similar teleconnections between tropical oscillations and mid-latitude weather, despite a separation of 40,000 km.  The perturbations observed in 2011-2014 could have been a unique consequence of the storm's timing, as we would expect the extent of the perturbation to be sensitive to the precise QPO phase, and it will be important to track the QPO behaviour during the next storm outbreak on Saturn.  Furthermore, even if the 1990 equatorial storm could have had a similar perturbing effect on the QPO (magnified as it occurred at a location geographically closer to the meteorological activity responsible for the QPO), this still cannot explain the mismatch of phase between the Voyager and Cassini observations [23] as the oscillation appeared well-behaved in the ground-based time-series during this period [3].  Conversely, the phase of the QPO may influence Saturn's extra-tropical circulation via secondary circulations, modulating temperatures at mid-latitudes and changing the atmospheric susceptibility to large-scale convective eruptions like that seen in 2010.  The key conclusion of this work is that a gas giant's equatorial oscillation can respond to meteorology elsewhere on the planet, just like Earth's QBO.  The same could be true of Jupiter's QQO, provided that waves spawned by jovian convective storms are able to interact with the QQO shear zones (i.e., that they can overcome confinement barriers in latitude and altitude).  Stratospheric waves over Jupiter's equatorial belts have been observed in relation to tropospheric plume events [30], but a comprehensive survey relating QQO changes to Jupiter's belt/zone upheavals is not yet available.  Nevertheless, this disruption to Saturn's QPO supports the view of giant planet atmospheres as intricately interconnected systems.

---

[*] Given that the magnitudes of stratospheric winds are subject to large uncertainties (see Methods), we cannot determine whether the stratospheric winds have switched to retrograde flow during this phase, as they do on Earth, so all winds are given relative to the 5-mbar level.

## Main References

**Materials and Correspondence**

Reprints and permissions information is available at www.nature.com/reprints. The authors declare no competing financial interests. Correspondence and requests for materials should be addressed to leigh.fletcher@leicester.ac.uk

**Acknowledgements**

LNF was supported by a Royal Society Research Fellowship and European Research Council Consolidator Grant (under the European Union's Horizon 2020 research and innovation programme, grant agreement No 723890) at the University of Leicester. The UK authors acknowledge the support of the Science and Technology Facilities Council. SG and TF were supported by the CNES. A portion of this work was completed by GSO at the Jet Propulsion Laboratory, California Institute of Technology, under contract with NASA. We are extremely grateful to all those Cassini team members involved in the planning, execution and reduction of the CIRS data, without whom this study would not have been possible. This investigation was partially based on VLT observations collected at the European Organisation for Astronomical Research in the Southern Hemisphere (see Extended Data for ESO program IDs); and on data acquired by the Subaru Telescope operated by the National Astronomical Observatory of Japan, and extracted from the SMOKA database (program IDs are provided in the Extended Data).

**Author Contributions**

LNF was responsible for analysing the nadir data and writing the article. SG and TF analysed Cassini limb observations and assisted with the nadir-limb comparison and calculation of zonal winds. LL provided a cross-comparison of zonal winds via a different algorithm, and MF provided assistance with the wind calculations. GSO assisted with the ground-based observing campaign. PGJI developed the software to permit inversions of Cassini/CIRS spectra. NG generated the CIRS spectral database. All authors read and commented on the manuscript.

**Competing Financial Interests**

The authors declare no competing financial interests.

**Figures:**

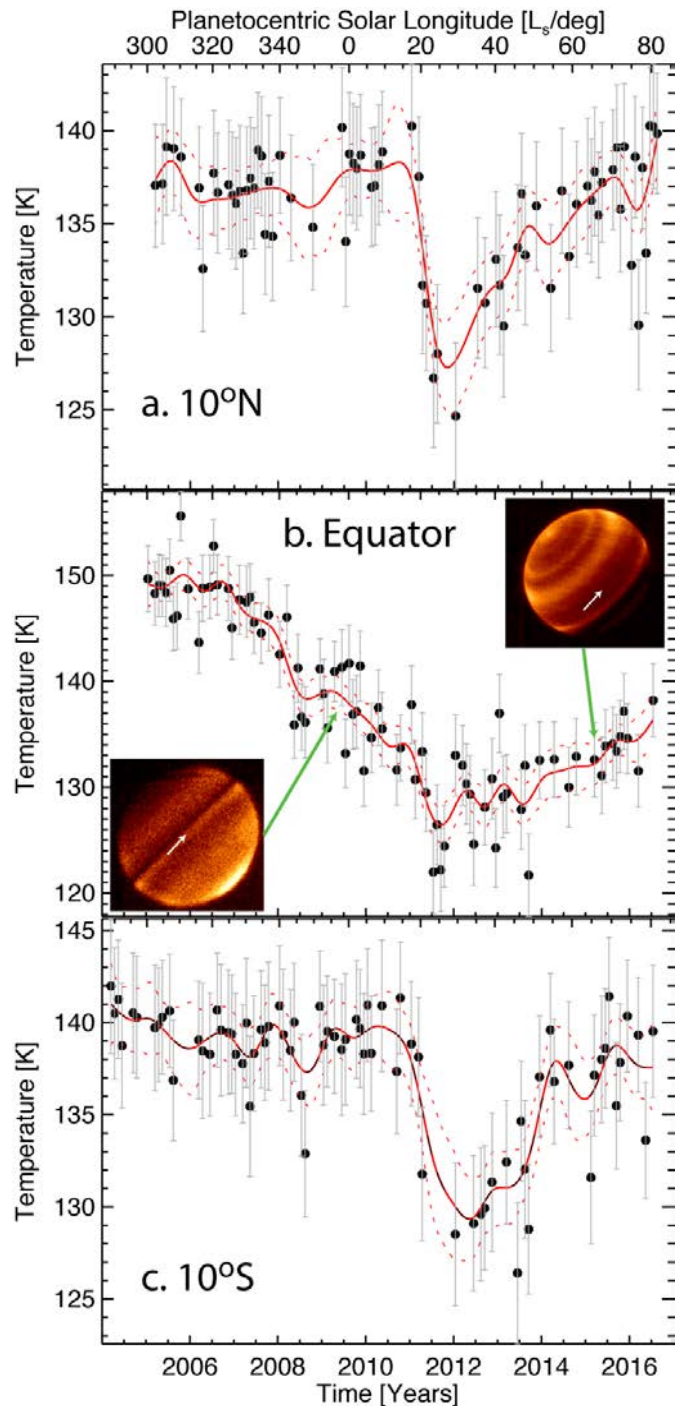

**Figure 1 Saturn's stratospheric temperatures at 1 mbar determined by Cassini/CIRS.** Tens of thousands of individual mid-infrared spectra were inverted to measure stratospheric temperatures (circles with retrieval uncertainties). Tensioned splines (solid lines with dotted uncertainties) are used to smoothly interpolate the time series. The dramatic cooling associated with the northern storm activity is evident both sides of the equator in panels a and c (the storm first erupted on December 5th 2010), with a return to 'pre-storm' temperatures in 2014. Inset images in b reveal 7.9-µm methane emission from the VISIR instrument on VLT on 2009-April-20 and 2015-May-21. The equator is shown via the white arrow. These confirm the disappearance of the bright equatorial band as the QPO switched phase (see Methods and Supplemental Information).

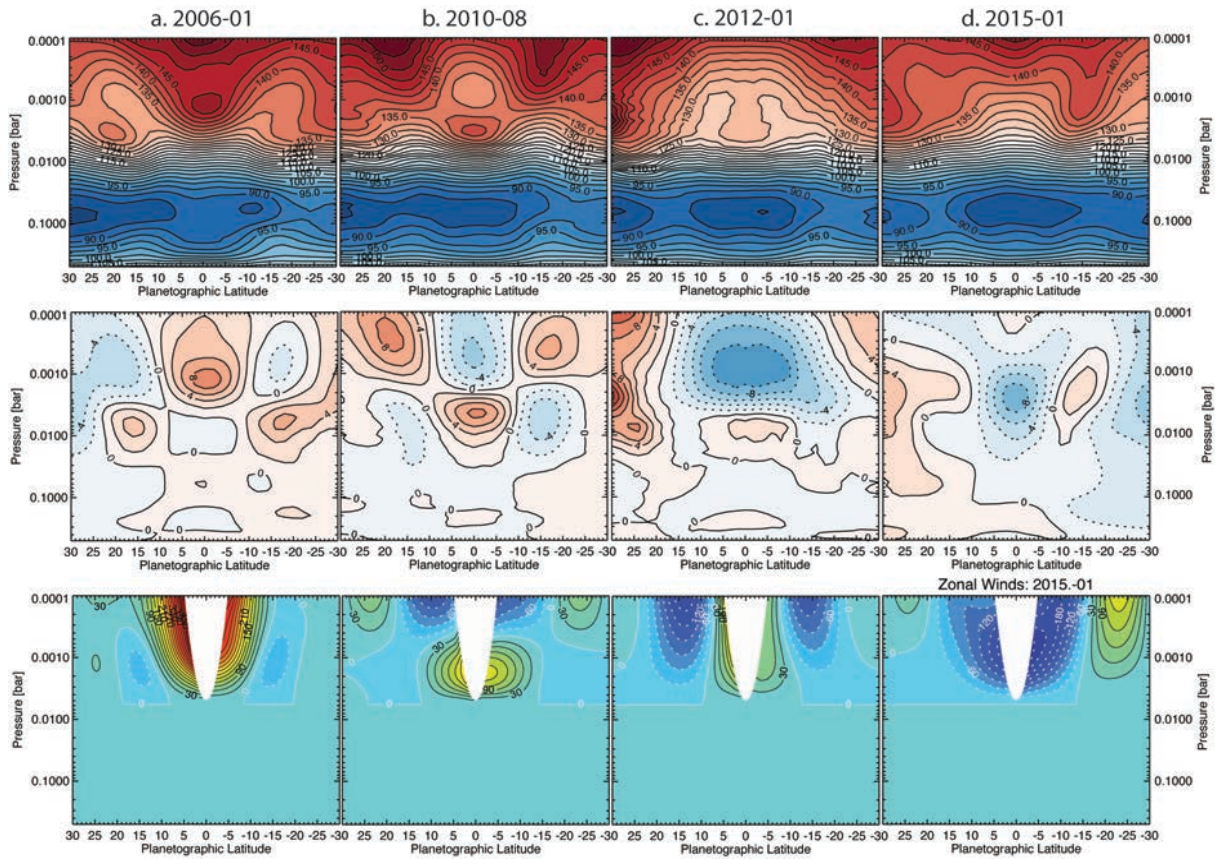

**Figure 2 Snapshots of Saturn's equatorial temperatures (top row) and winds (bottom row) during the 13-year time-series in Fig. 1** (see animations in Supplemental Information). The temperature anomaly in the middle row was formed by subtracting a time-averaged temperature profile for each latitude from the data in the top row (excluding 2011-2013). Extra-tropical signatures of the QPO (in antiphase with the QPO structure) can clearly be seen at ±15-20° in both hemispheres. Zonal winds in the bottom row were calculated from the modified thermal windshear equation, integrating along cylinders parallel to Saturn's rotation axis [18, 19] from a level-of-no-motion at 5 mbar, and therefore should be considered as a perturbation on top of the middle-atmospheric winds which are likely to remain eastward at these altitudes (see Methods). Our results do not imply retrograde stratospheric flow. The white parabola indicates regions where the thermal wind calculation is unconstrained [21]. Temperature uncertainties are 2-4 K, but zonal-wind uncertainties are large given the low vertical resolution (see Methods). Temperatures and winds are shown during Cassini's prime mission (2006), before the storm onset (2010), at the maximum equatorial cooling (January 2012), and once the QPO had returned to the expected westward phase (with a cool band at the equator at 1 mbar) in 2015. Full animations of these figures can be found in the Supplemental Information.

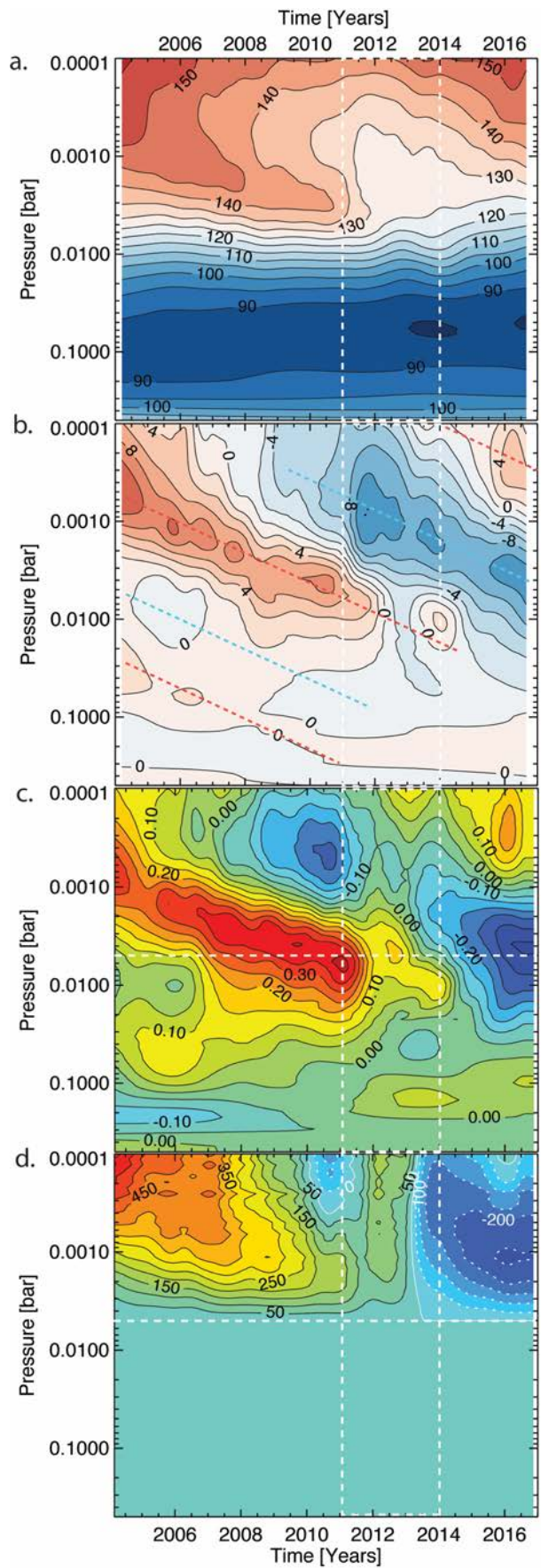

**Figure 3 Hovmöller diagrams showing the downward propagation of the QPO thermal and wind patterns and the 2011-14 disruption.** Panels show (a) equatorial temperatures [K]; (b) temperature anomalies [K]; (c) thermal windshear along cylinders parallel to Saturn's rotation axis (m/s/km) estimated from the modified thermal wind equation (TWE, see Methods); and (d) zonal winds (m/s) in the p<5 mbar region estimated by integrating the TWE upwards from *u=0* m/s at 5 mbar (i.e., only in the altitude region to which nadir data are sensitive, signified by the horizontal white dashed line in c, d). The vertical dashed lines in 2011 and 2014 indicate the lifetime of the stratospheric vortex near 40°N [10]. All values were averaged over ±5° of the equator, although we caution that nadir inversions (i) lack information about the latitudinal temperature gradient in the 5-to-80 mbar range; (ii) lack vertical temperature resolution which causes significant uncertainty on the magnitude of thermal winds (see Methods). Westward winds (relative to the 5-mbar level) are shown by dashed contours in (d). Diagonal dotted lines in (b) show the positive and negative peaks of the QPO pattern.

**Methods**

**Cassini/CIRS Spectra**
This study employs Cassini/CIRS [24] interferograms of Saturn taken between March 2004 and November 2016. We use 15-cm$^{-1}$ spectral resolution mid-infrared observations spanning 600-1400 cm$^{-1}$ (7.1-16.6 µm) acquired using CIRS focal planes 3 and 4, where each pixel covers a 0.273 mrad$^2$ instantaneous field of view. In addition, we truncated longer CIRS interferograms that had been designed to produce higher spectral resolutions (2.5 cm$^{-1}$) to expand the 15-cm$^{-1}$ dataset. Interferograms were calibrated and converted to spectra that were coadded on a 1-month time grid and a 2$^o$-wide latitudinal grid (with 1$^o$ step size). Both prime CIRS-targeted observations, and observations 'riding along' with other Cassini instruments, are employed to expand the time series, following techniques described by Fletcher et al. [31]. This results in the temporal coverage of Saturn's low latitudes as shown in Fig. S1 of the Supplemental Information. Note that the spectra are not truly zonally-averaged, as CIRS did not acquire full 360$^o$ longitude coverage every month. However, away from Saturn's storm, the planet's temperature structure is nearly axially symmetric, such that the coadded spectra represent good approximations to the zonal mean (note that this is not true of latitudes 30-60$^o$N during the 2011-14 epoch, which is why these latitudes are omitted from the main article). Furthermore, we assessed all CIRS maps that did cover 360$^o$ of longitude during the storm epoch (2011-2013). Although horizontal wave activity was present, we did not detect significant longitudinally-localised perturbations in the temperatures at low latitudes. We conclude that the wave-induced momentum changes acted at all longitudes.

**Spectral Inversion and Information Content**
Cassini CIRS spectra of Saturn are characterised by a host of stratospheric emission features (e.g., ethane, acetylene, and methane) and tropospheric absorption features (e.g., phosphine) superimposed onto the broad collision-induced absorption spectrum of $H_2$-$H_2$ and $H_2$-He. These are inverted using an optimal estimation retrieval algorithm, NEMESIS [32], based on the formalism of optimal estimation developed by the Earth-observing community [33]. Inversions balance the goodness-of-fit to the spectral data against constraints imposed by *a priori* information – specifically, the vertical temperature and gaseous profiles provided by previous studies (see Fletcher et al. [10] for full details of the inversion procedure, sources of spectral line data and *prior* distributions, and the conversion of line data to *k*-distributions for rapid calculation of atmospheric transmission). Uncertainties on the retrieved temperatures depend on measurement errors, uncertainties in spectral line data, assumptions about vertical smoothing in the retrieval process, and correlations between parameters. Temperature uncertainties range from ~3 K at the tropopause, to 1.5-2.0 K near 200 mbar (the peak of the tropospheric contribution), 3.5-4.5 K at 1-5 mbar (the peak of the stratospheric contribution). The same time-independent and latitude-independent *a priori* initial profile was used for all inversions, based on a combination of low-latitude nadir and limb inversions from previous studies [20,21], so as not to introduce biases in the results. Retrieved temperatures representative of the zonal mean at 0.5, 1.0 and 5.0 mbar are shown in Fig. S2 of the Supplemental Information. The temperatures at individual latitudes and pressures were then interpolated (Fig. S3) to produce Fig. 1 of the main article. The temporal evolution of these interpolated temperatures, and their associated temperature anomalies, is shown in Fig. S4 for a range of latitudes.

This study utilises nadir data only (defined in this context as any observation with an emission angle less than 60°), and so vertical information content is low, particularly in the middle atmosphere. Temperature gradients can only be reliably extracted in the 0.5-5.0 mbar range, and even then, they would underestimate any sharp vertical gradients in temperature. The nadir 600-1400 cm$^{-1}$ data have no sensitivity to temperatures in the 5-80 mbar range, meaning that latitudinal thermal gradients ($\frac{\partial T}{\partial y}$) will not be accurately determined from nadir data alone. Figures S5 and S6 of the Supplemental Information compare nadir spectral inversions at 0.5, 2.5 and 15.0 cm$^{-1}$ to CIRS limb observations previously published [3, 21]. Higher spectral resolutions better resolve the shapes of the spectral lines to extract more vertical information. Allowing for differences in the observing strategies, the comparison between the two independent retrieval methods is favourable in the 0.5-5.0 mbar range. Some discrepancies do exist that are outside of our formal retrieval uncertainties in the 0.5-5.0 mbar range, but these are likely due to incomplete spatial sampling on Saturn, particularly during the storm epoch.

Offsets between the two data types (nadir and limb) are larger at 10 mbar, where only the limb data offer any constraint. Nadir sensitivity is dramatically reduced at these pressures as the Planck function of the CH$_4$ emission drops sharply due to the cooler temperatures for p>5 mbar. This poses a significant problem for stratospheric wind-speed estimates via the thermal wind equation, as we describe below. As a cross-check that the QPO perturbation is a real phenomenon in Saturn's atmosphere, Fig. S7 presents the same time series as Fig. S3, but using only 2.5-cm$^{-1}$ observations at their native spectral resolution. These have poorer temporal sampling, but show the same trends as the 15-cm$^{-1}$ dataset.

**Thermal Winds**
Estimation of stratospheric zonal winds ($u$) rely on integrations of the cloud-tracked zonal winds using the measured temperature gradients, $\frac{\partial T}{\partial y}$. The standard thermal wind equation (TWE) relates zonal vertical windshears to meridional temperature gradients in pressure $\left( \frac{\partial u}{\partial z} = -\frac{g}{fT} \frac{\partial T}{\partial y} \right)$, where $z$ is the altitude, $f$ is the Coriolis parameter ($f = 2\Omega \sin \phi$ where $\phi$ is the planetographic latitude and $\Omega$ the planetary rotation rate), and $g$ is the local acceleration of gravity. This equation breaks down at low latitudes as the Coriolis parameter tends towards zero. Geostrophic balance can still apply at the equator, but computation of zonal winds is impracticable due to error amplification in altitude owing to the small value of $f$. Away from the equator, and for thin atmospheres, the wind gradient is vertical. However, Flasar et al. [19] showed that the integration of the TWE near the equator must be along cylinders concentric with the rotation axis ($z_{||}$, rather than with altitude, as in the standard TWE). We estimated the zonal winds using this modified TWE using two different approaches [18, 21] and found both to be consistent with one another. Following Guerlet et al. [21], the modified TWE can be expressed as:

$$\frac{\partial}{\partial z_{||}} \left( 2\Omega u + \frac{u^2}{r \cos \varphi_c} \right) = -\frac{g}{T} \frac{1}{r} \left( \frac{\partial T}{\partial \varphi_p} \right)_P$$

where $r$ is the local radius, $\varphi_c$ and $\varphi_p$ are planetocentric and planetographic latitudes, respectively, and all other terms have been defined previously. We display temporal trends in the thermal windshear along these cylinders in Fig. 3c and the Supplemental Information Figure S8. Furthermore, Figure S9 demonstrates that the vertical windshear (from the standard TWE, omitting low latitudes where errors are magnified) and the windshear along cylinders (from the modified TWE) are morphologically similar, but the latter is the correct form to use at the equator.

Integration of the modified TWE requires a selection of boundary conditions. We can choose to (i) use the cloud-top winds from continuum-band imaging [26] as our boundary condition at 500 mbar, following previous wind calculations [18,27]; or (ii) specify an artificial level of zero motion ($u=0$ m/s) at an altitude just below our region of vertical sensitivity, following Fouchet et al. [4]. Note that we do not expect a zero-wind level in the 5-70 mbar range, as Cassini cloud tracking in methane-band imaging [34] suggests that the winds are still strong and eastward at the 50-mbar level. The winds presented in the p<5 mbar range in Figs. 2 and 3 should therefore be considered as a 'delta' on top of the background flow. Furthermore, the modified TWE can only be integrated for latitudes where the tangent cylinders intersect the reference pressure $p_0$ where the boundary condition is specified. If the tangent cylinder intersects lower pressures, then no boundary condition is available for the wind integration (i.e., we do not know the zonal velocity in the equatorial plane for $p<p_0$ mbar). This 'no-solution' region is defined by a parabola, centred on the equator, in Figure 2 and the Supplemental Information. In producing the temporal plots in Fig.3, we extrapolated through the 'no-solution' region to the equator so that the values represent those calculated at the edge of the no-solution parabola.

Fig. S10 contrasts the wind calculations for zero-motion levels at 5, 10 and 20 mbar, with a full calculation assuming the Cassini cloud-tracked winds at 500 mbar. Although the latter matches calculations from Li et al. [18, 34], the wind speeds are deemed to be unrealistically large for two reasons: (i) the nadir data do not adequately constrain meridional temperature gradients in the 5-to-80 mbar range; and (ii) the vertical resolution of nadir inversions is sufficiently poor that regions of windshear are smeared out over broad altitude ranges. As limb inversions have shown that $\frac{\partial T}{\partial y}$ at 10 mbar is generally opposite to that at 1 mbar (Fig. S5), a nadir-only wind calculation will be severely biased to the $\frac{\partial T}{\partial y}$ measured in the 1-mbar region, leading to a substantial overestimation of winds there. This explains why the magnitude of the 1-mbar jet from nadir inversions [18, 27] is significantly larger than that from limb inversions [4, 21]. CIRS limb analyses suggest a ~300 m/s difference between zonal winds in the eastward and westward shear zones [21]. Fig. 3d uses a zero-motion constraint at 5 mbar and finds a ~600 m/s difference from 2005 to 2016, so we estimate that nadir-derived zonal-wind contrasts are subject to uncertainty factors of approximately 2-3. However, placement of the zero-motion boundary at higher pressures (Fig. S10) makes this contrast even larger, highlighting the uncertainty in absolute wind measurements.

The estimated windshear and zonal winds are compared in Figure S8 for both scenarios (i.e., integrating only for p<5 mbar, and integrating for p<500 mbar). The latter suggests that stratospheric winds turn retrograde in the westward QPO phase, but we consider this unlikely give the strong prograde flow at Saturn's equator at the cloud tops, and is an artefact of our lack of knowledge of the $\frac{\partial T}{\partial y}$ structure in the lower stratosphere. Although Cassini limb observations could help to resolve this conundrum, they are too infrequent to develop a proper time series of the varying 10-50 mbar temperature gradients, so only nadir data can reveal the middle-atmospheric changes associated with this storm over short timescales. This remains a fundamental limitation of determining winds from nadir spectroscopy at low spectral resolutions.

**Ground-Based Thermal Observations**
In addition to Cassini observations of the vertical structure of the equatorial oscillation, Saturn's QPO was also discovered in the thermal contrasts displayed in ground-based imaging sensitive to stratospheric methane (7.9 μm) and ethane (12.3 μm) [3]. This imaging programme, using the Subaru Telescope and Very Large Telescope, has continued throughout the Cassini mission, and proved instrumental in tracking the evolution of the 2010 storm and its stratospheric vortex [10]. Using the CIRS time series generated from the spline interpolation of the inversions, we forward-modelled the expected radiances in typical 7.9- and 12.3-μm filters, taking care to use the correct observational geometry for Earth-based facilities. The effects of the 2011-13 equatorial cooling can clearly be seen in both filters in Fig. S11 of the Supplemental Information, and these data confirm the change in the QPO phase. Furthermore, if we compute the same contrast-index as Orton et al. [3] – namely, the difference between 3.5°N and 15.5°N (red lines) and 3.5°S and 15.5°S (blue lines, all latitudes planetographic) – we can compare the simulation to observations. We consider data from (a) the VLT VISIR instrument [35], from 2008 (program ID: 381.C-0560), 2009 (383.C-0164), 2010 (084.C-0193) and 2011 (287.C-5032), and then more recently in 2015 (095.C-0142) and 2016 (097.C-0226); and (b) the Subaru COMICS instrument [36,37] from 2005 to 2009 in programs S05A-029, S07B-015, S07B-076, and S08B-023. Subsets of the VISIR 7.9-μm observations are shown in Fig. 1 and Fig. S12. Data were reduced, remapped and calibrated using techniques described in Fletcher et al., [38]. The equatorial region (±3.5°) was warmer than the tropical latitudes (±15.5°) at 7.9 μm until 2010/2011. After this time, the whole ±20° latitude range grew cooler, and the contrast between the equator and tropical latitudes reversed. By the end of the sequence, the warm equatorial band observed by Orton et al. [3] had been replaced by a cool equatorial band. Unfortunately, the data in Fig. S11 lack the time coverage to add more detail to the rapid changes in 2011. Furthermore, images in 2011 were dominated by the intense brightness of the stratospheric vortex, making it difficult to observe the equatorial contrasts from Earth. This Earth-based program will continue beyond the end of Cassini's mission in 2017 to track the expected return of the warm equatorial band in 7.9-μm imaging, which can be seen forming at low pressure (p<0.5 mbar) at the end of the Cassini time-series.

**Data Availability**

This work relies on two data sources – Cassini and ground-based data. Calibrated Cassini/CIRS spectra and raw interferograms are available via the Planetary Data System (http://atmos.nmsu.edu/pdsd/archive/data/co-s-cirs-234-tsdr-v32/). Coadded spectra, retrieved atmospheric temperatures, spline-interpolated temperatures and other derived products are available from the primary author. Raw VLT/VISIR images from 2008-2016 are available via the archive of the European Southern Observatory (ESO, http://archive.eso.org). Raw Subaru/COMICS images from 2005-2009 are available via the SMOKA Science Archive (http://smoka.nao.ac.jp/). Calibrated images and maps are available from the primary author.

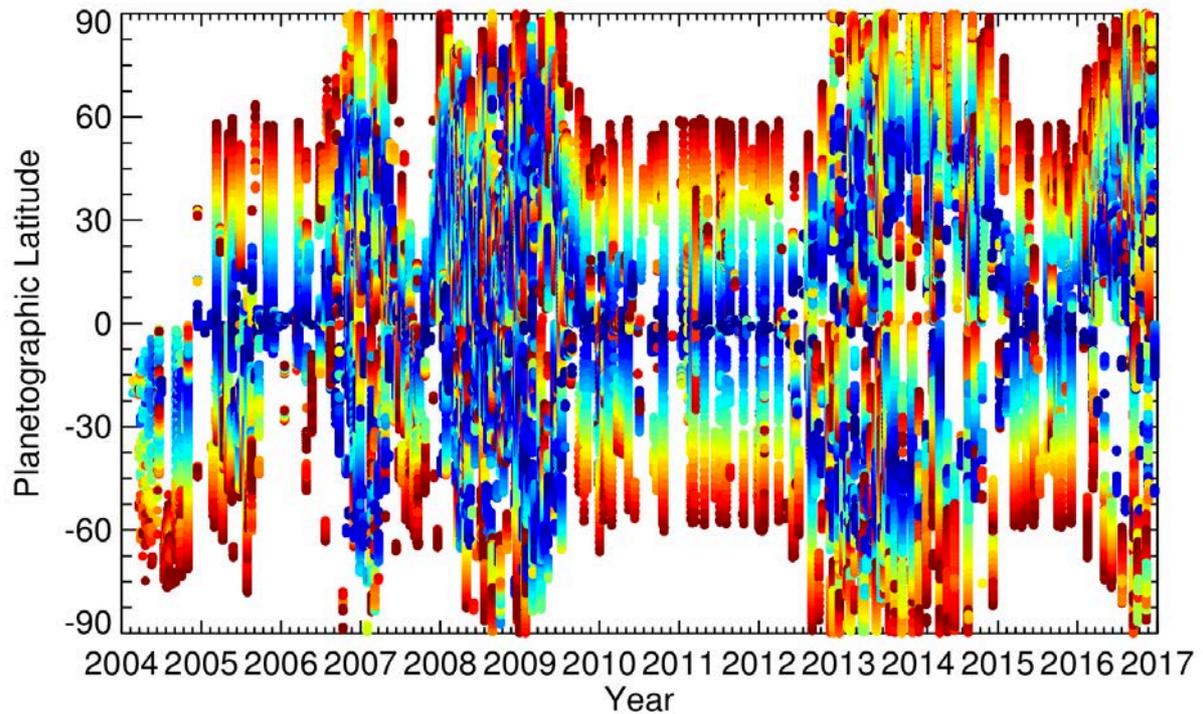

**Supplemental Figure 1:** Spatial coverage of Cassini/CIRS spectra used in this study. Each point represents a spectrum averaged over multiple longitudes in a particular month. Colours denote observation zenith angles, from blue ($0^o$) to red ($60^o$), the maximum used in this study. Cassini's inclined orbital periods can clearly be seen, providing access to higher latitudes.

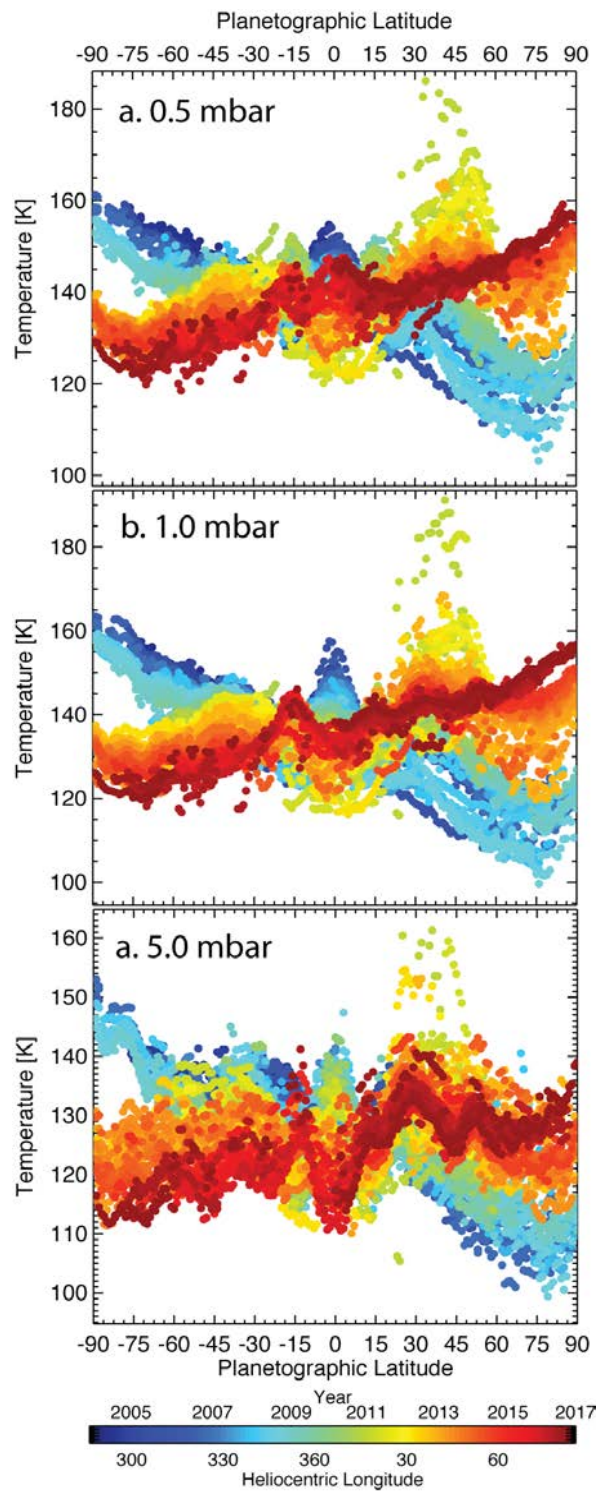

**Supplemental Figure 2:** Zonal mean temperatures in Saturn's stratosphere retrieved from the CIRS time series. Trends at a particular latitude and pressure are shown in Fig. 1 and **Supplemental Figure 3**. Uncertainties are 3-4 K at these pressure levels (see **Supplemental Figure 3**), depending on the number of spectra contributing to a particular average. The effects of the northern beacon can be seen from 30-50°N in 2011-13. Results are coloured according to the date of observation, following the key below the figure. This study focusses on the changing contrasts at Saturn's equator, and the results at 60-90° in both hemispheres have been published [31].

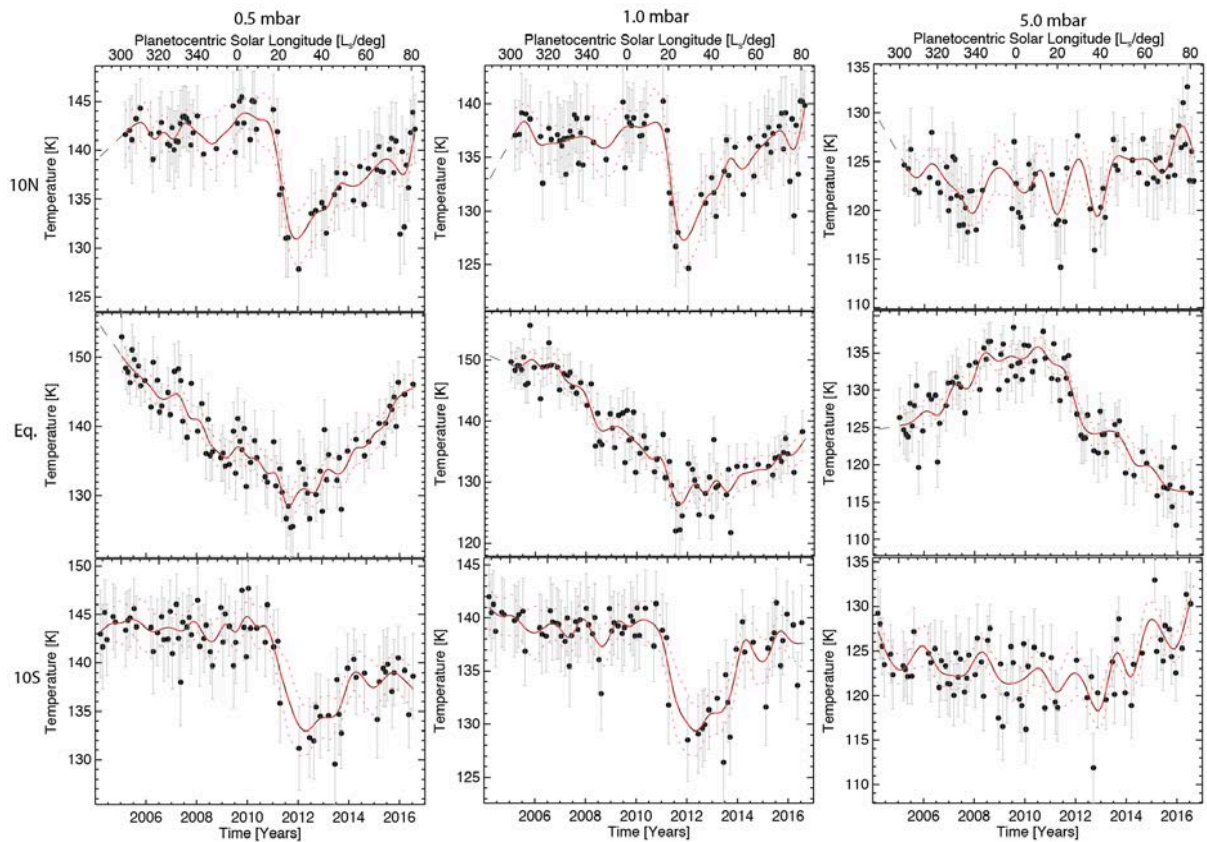

**Supplemental Figure 3**: Temperature trends at 0.5, 1.0 and 5.0 mbar for ±10° and the equatorial zone. The central column is used as Fig. 1 of the main article. Temperature uncertainties for each point are shown as grey vertical bars. The red curves are tensioned splines fitted through the time series [39]. The disruption from the 2011 storm can only be seen for pressures less than 5 mbar, with the 5-mbar time series showing trends consistent with that expected for Saturn's QPO. The interpolated time series were used to generate the temperature/wind snapshots in Fig. 2 and the Hovmöller diagrams in Fig. 3.

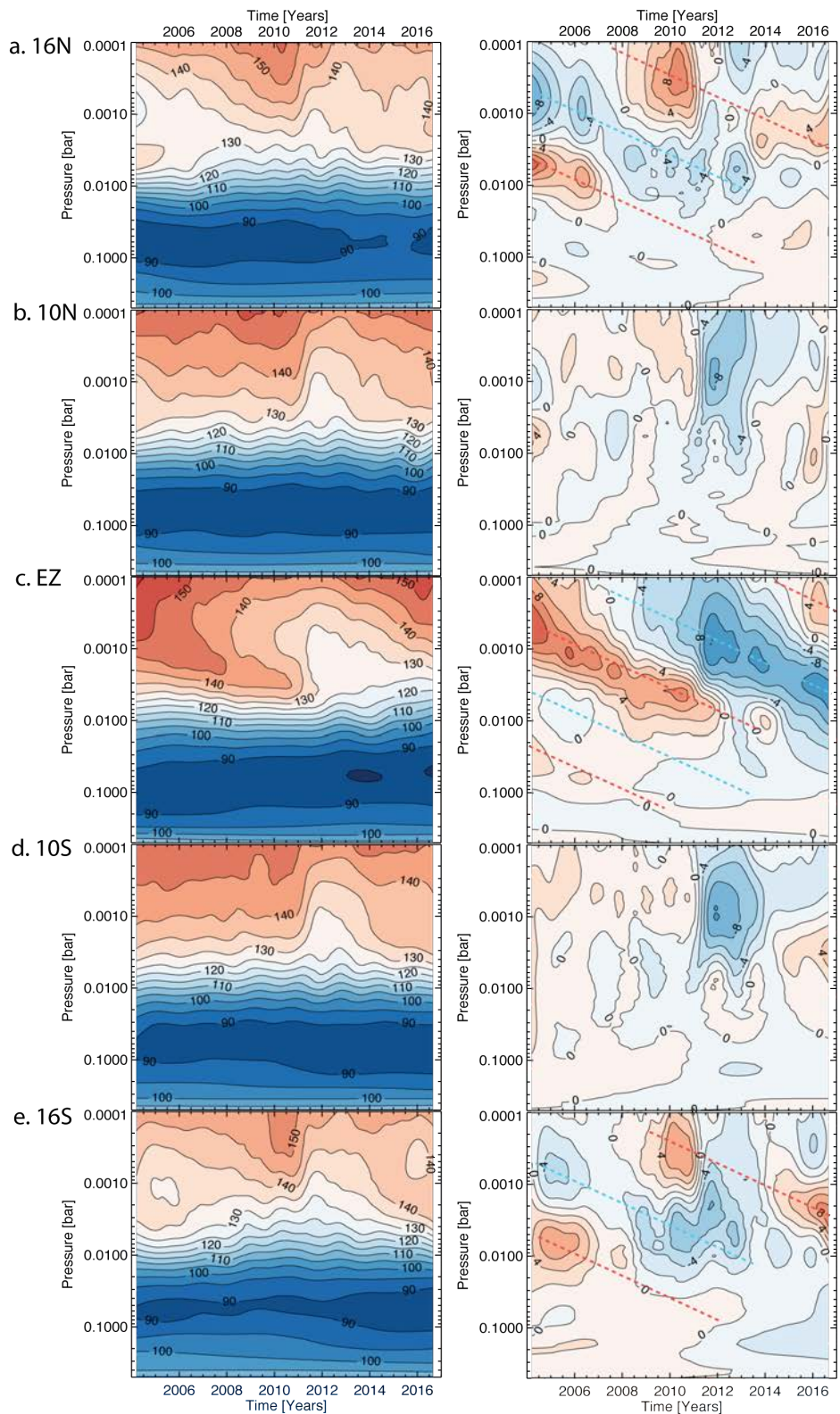

**Supplemental Figure 4:** Hovmöller diagrams showing temperatures (left) and temperature anomalies (right) averaged over ±2° latitude at ±16°, ±10° and the equator. Anomalies were estimated by subtracting the 2004-2017 average temperature profile from each latitude (excluding dates between 2011-2013). Diagonal dashed lines in red (warm) and blue (cool) show the downward propagation of these opposite QPO phases, as well as their extratropical counterparts. The QPO disruption is clearly seen at ±10° as a substantial cooling for a 2-3 year period.

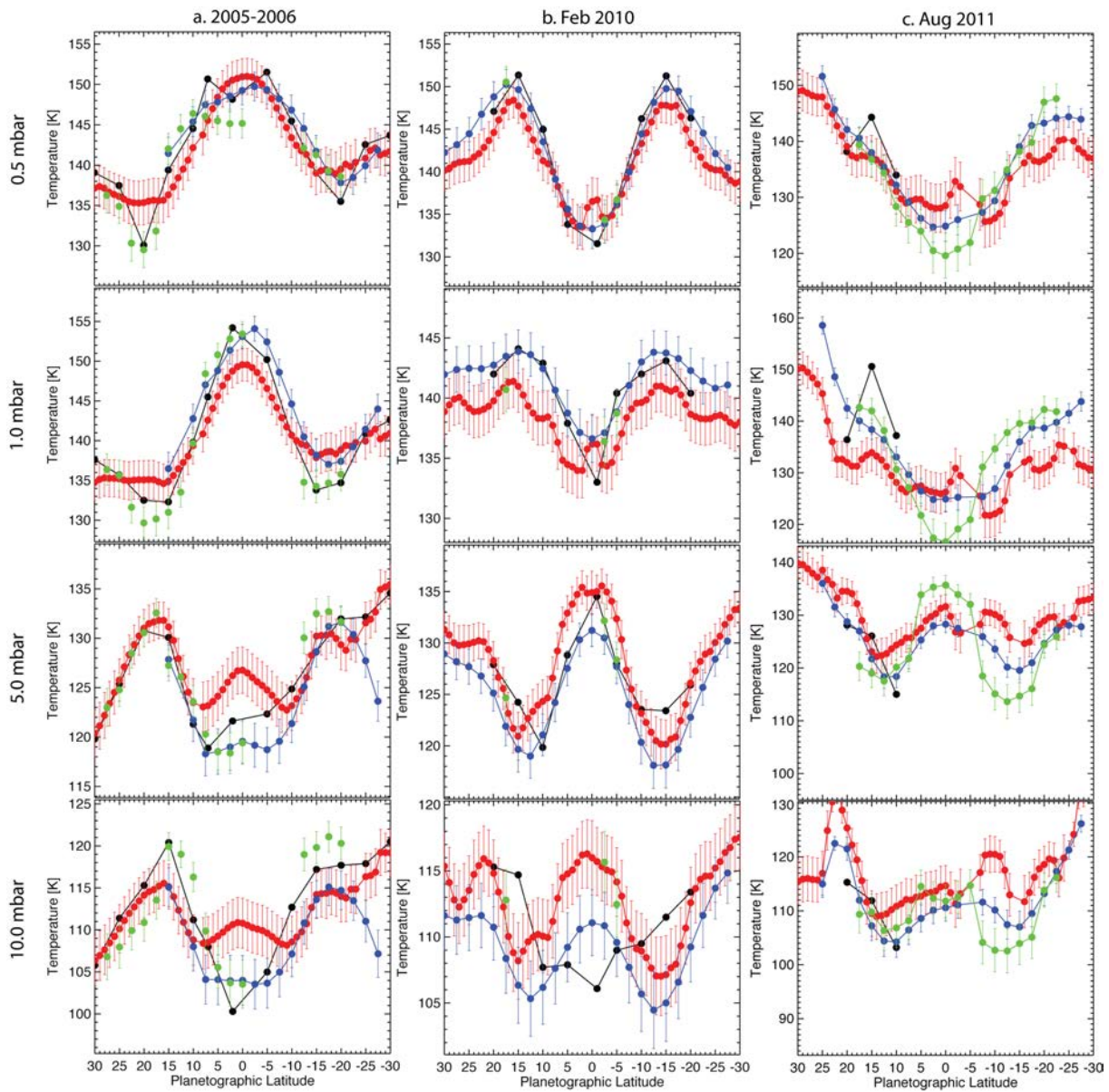

**Supplemental Figure 5:** Comparison of nadir retrievals to limb measurements from Fouchet et al., [4] and Guerlet et al. [9] (black points). Although the observations are not perfectly coincident and sometimes cover different regions of Saturn's atmosphere, the comparison between absolute temperatures between the 0.5-10 mbar range is excellent, despite the use of two entirely independent retrievals methodologies. Inversions are shown using three different nadir spectral resolutions: 0.5 cm[-1] (green), 2.5 cm[-1] (blue) and 15.0 cm[-1] (red). The comparison is poorest at 10 mbar, where the nadir data lack sensitivity. Retrievals from 15 cm[-1] tend to under-predict temperature contrasts compared to the higher spectral resolutions. However, as the temporal coverage of the 15-cm[-1] data is better than any other data types, it is used throughout the main article.

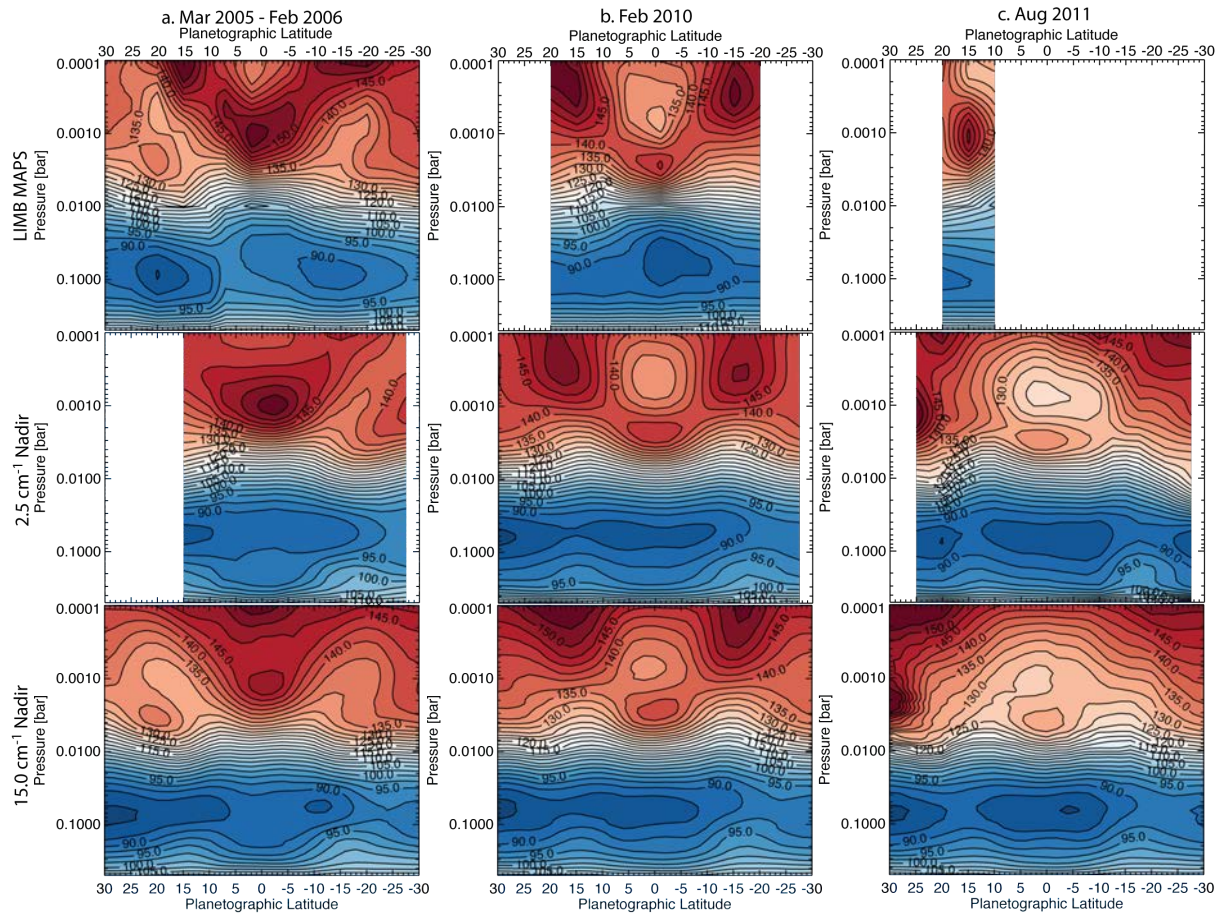

**Supplemental Figure 6:** Comparison of zonal mean temperatures derived from Cassini/CIRS limb data (top row), 2.5-cm⁻¹ resolution nadir data (middle row) and 15.0-cm⁻¹ resolution nadir data (bottom row). The limb observations in 2005-06 and 2010 have been previously published [4,9]. Although the limb data has the best vertical information content and is sensitive to p<30 mbar, which is partially matched by the high-resolution nadir data, we must rely on the low-resolution nadir data to provide the best spatial and temporal coverage of the stratospheric oscillation. This figure demonstrates that the qualitative structure is similar between each technique, although quantitative differences in absolute temperatures are evident, particularly in the p<0.5 mbar and 5-80 mbar range where nadir observations lack vertical sensitivity. Note that January 2006 was chosen as a comparison to the 2005-06 limb observations.

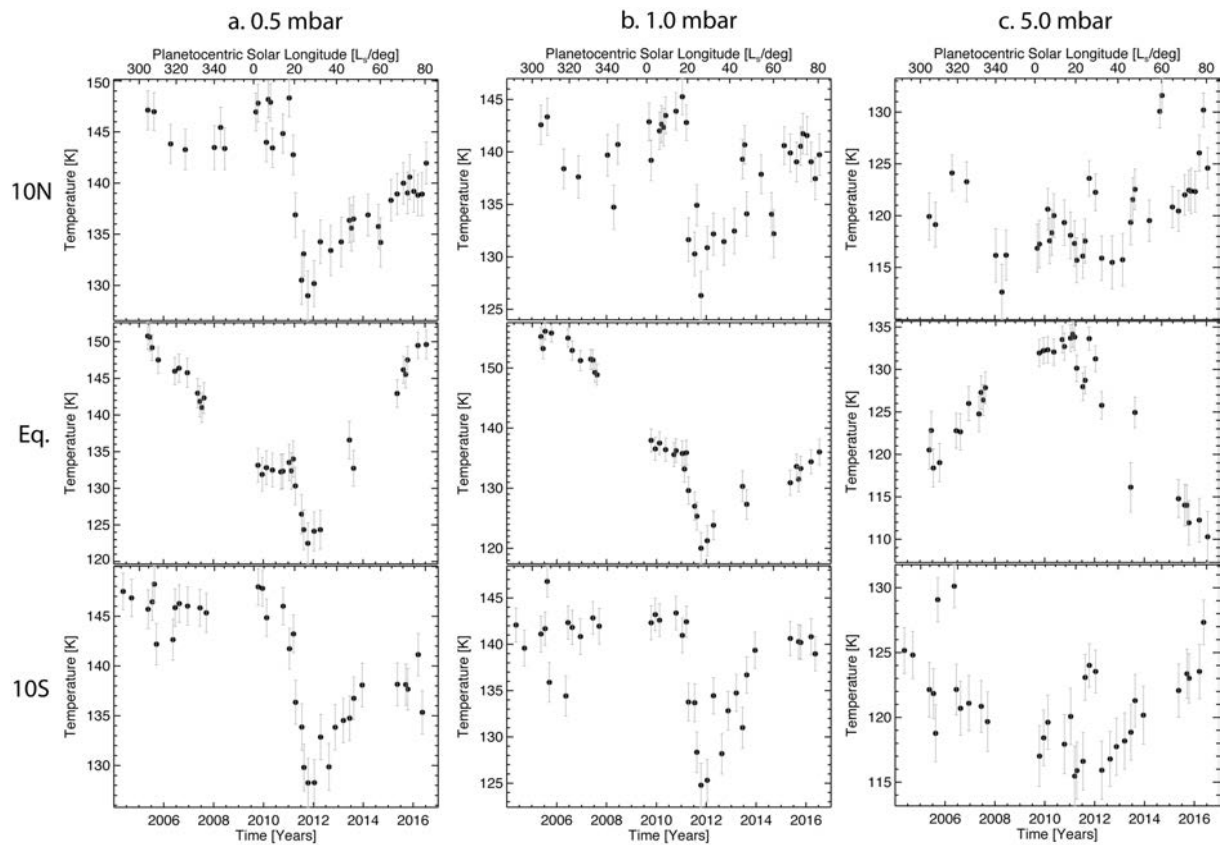

**Supplemental Figure 7:** Time series of 0.5, 1.0 and 5.0-mbar temperatures when we consider the 2.5-cm$^{-1}$ CIRS nadir data at their native resolution. This increases the vertical information content at the expense of coarser temporal coverage. Results show more scatter and are generally noisier (particularly at 5 mbar). However, the same 2011-13 cooling is evident at 0.5 and 1.0 mbar, with no sign of the perturbation at 5 mbar. These also suggest that the 15-cm$^{-1}$ inversions underestimate the cooling right at the equator, suggesting that the results reported in the main text are conservative.

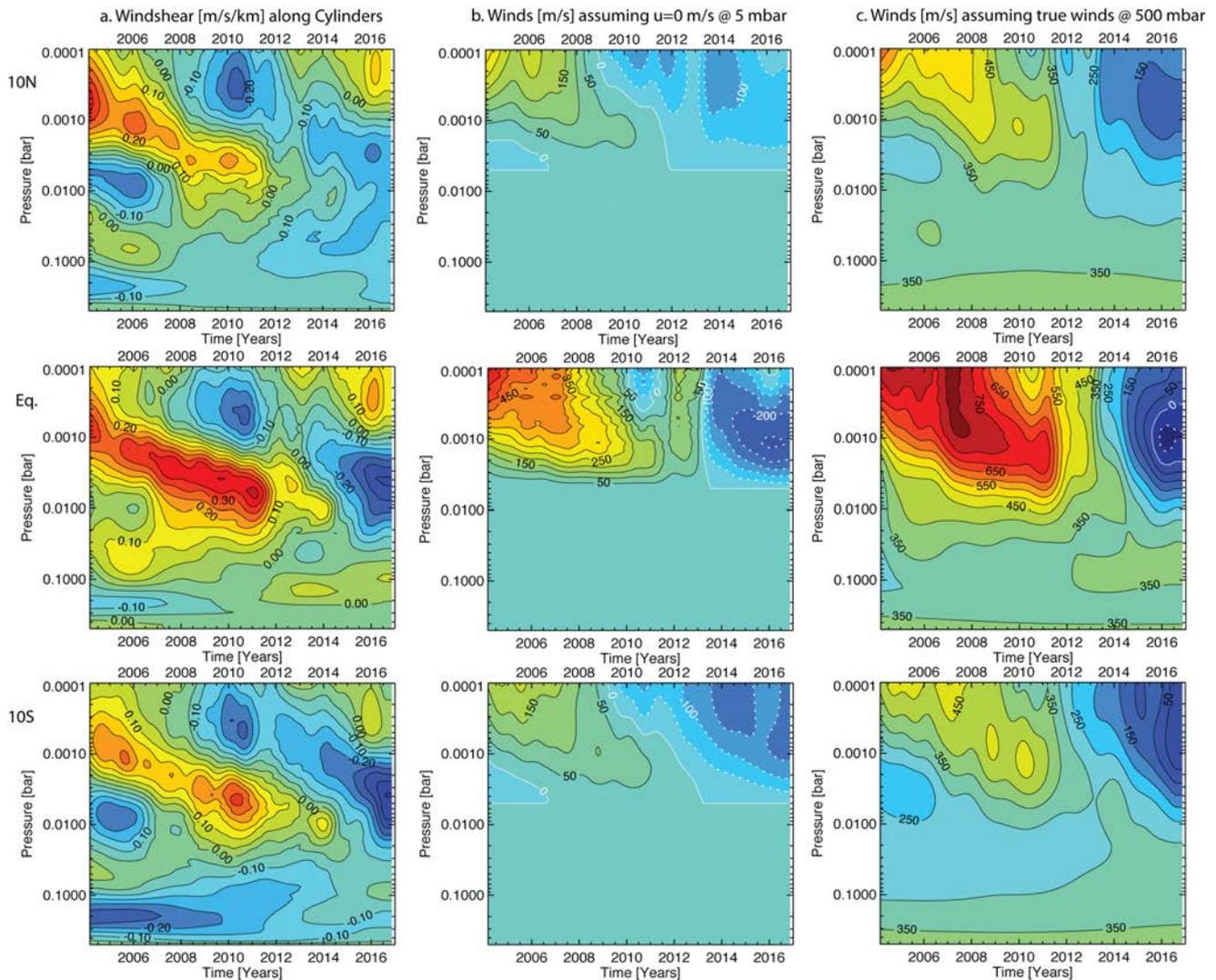

**Supplemental Figure 8.** Hovmöller diagrams for 10°N, the equator and 10°S, showing (a) the vertical windshear along cylinders in m/s/km, and the zonal winds calculated from these shears using the technique of [4,9] using two assumptions: (b) integrating from a level of zero motion at 5 mbar; and (c) integrating from 500 mbar upwards using the cloud-tracked winds [26] as the boundary condition. White, dashed contours are shown for negative wind values. Systematic uncertainties on the absolute zonal winds are very large (O(100 m/s)) due to the absence of vertical resolution and dT/dy constraint in the 5-80 mbar range, so that column (b) is a better representation of expected winds in the stratosphere. The downward propagation of alternating wind regimes, as well as the 2011-2014 disruption, are observable in this figure.

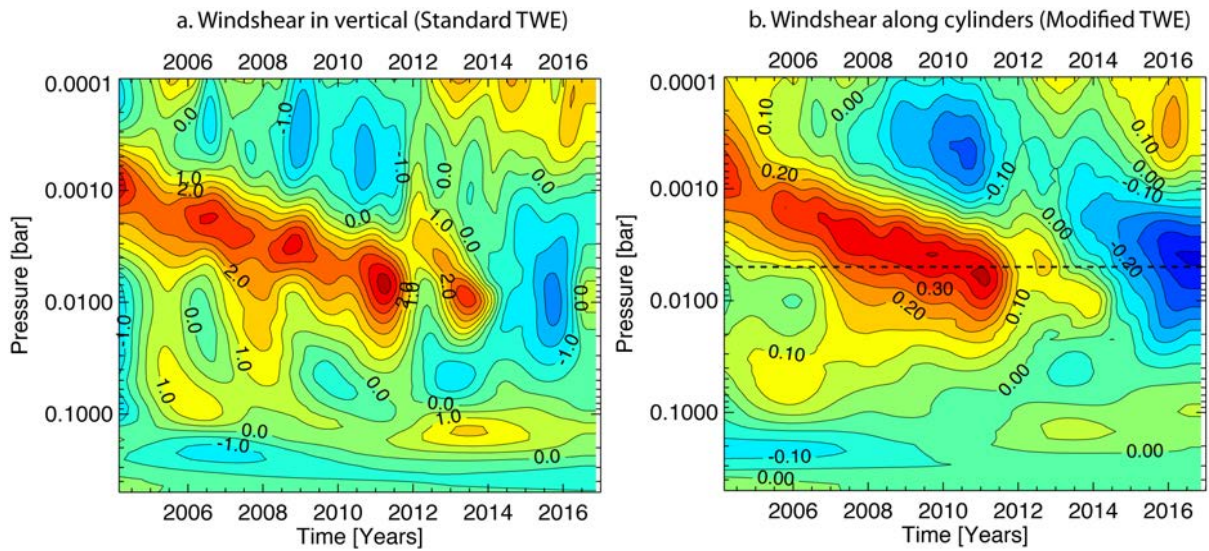

**Supplemental Figure 9.** Comparison of the temporal trends in the thermal windshear (m/s/km) when derived (a) in the vertical from the standard thermal wind equation (TWE); and (b) along cylinders parallel to Saturn's rotation axis in the modified TWE. Both are morphologically similar but differ in magnitude because of the different directions in which the windshear is measured. The horizontal dashed line in (b) shows the zero-motion boundary used in the main article.

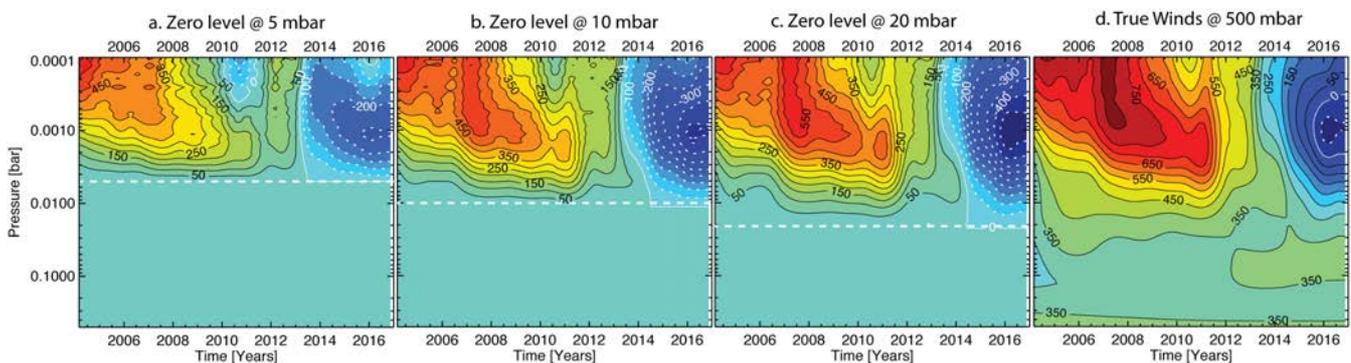

**Supplemental Figure 10:** Comparison of zonal wind calculations (using the modified TWE) assuming different boundary conditions, demonstrating uncertainty on the absolute magnitudes of the calculated winds. Placing the boundary at higher pressures (panels a-c) implies integrating over altitude domains where windshear information is less certain, and has the effect of increasing the calculated thermal winds by ~200 m/s (compared to placing the level-of-no-motion at 5 mbar). Panel d shows the equatorial winds if we use cloud-tracked Cassini measurements [26] as the boundary condition at 500 mbar, achieving winds of a similar magnitude to Li et al., [19].

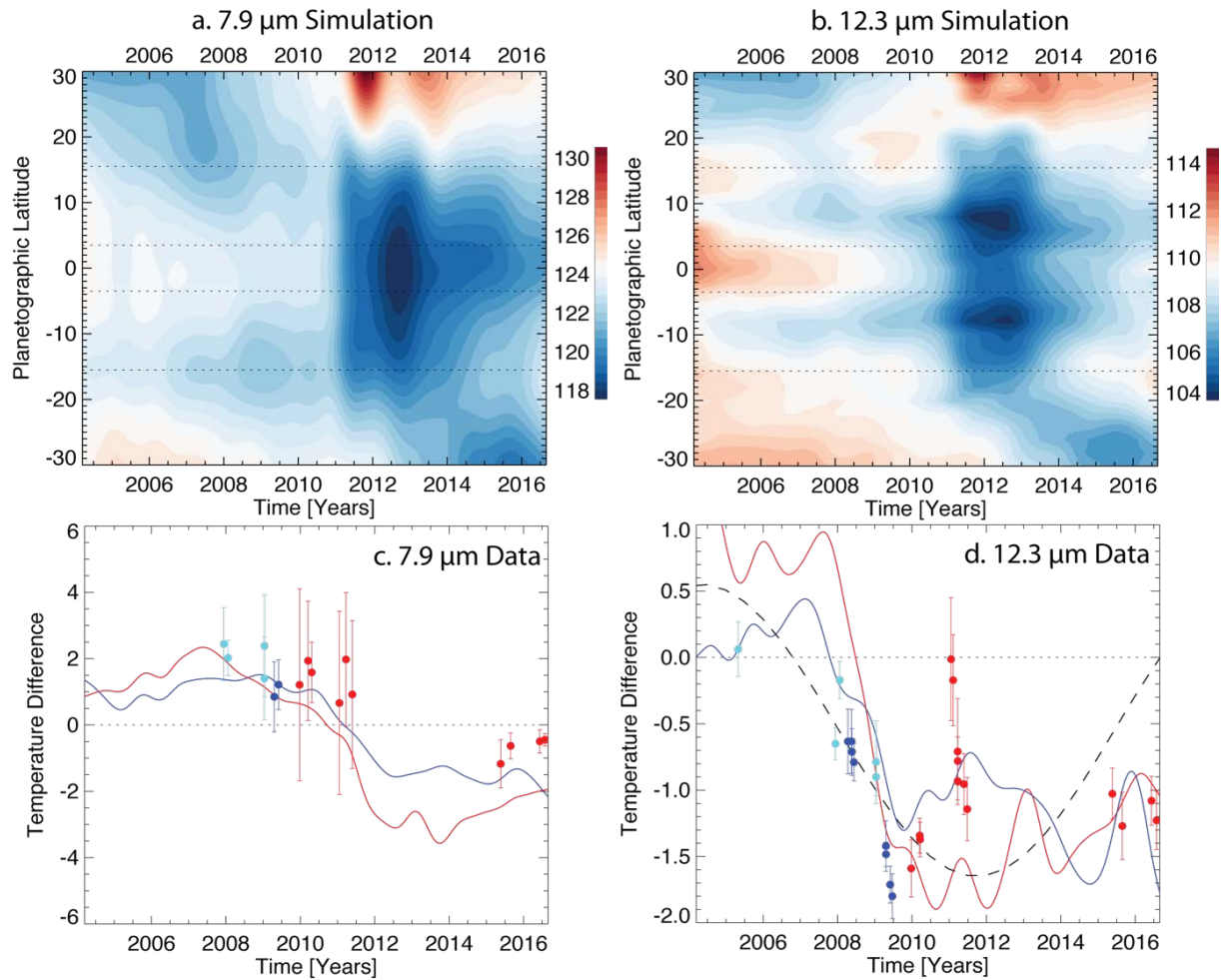

**Supplemental Figure 11:** Simulated appearance of Saturn's low latitudes in ground-based observations from telescopes with 8-m primary mirrors (e.g., VLT and Subaru). Zonally-averaged brightness temperatures are shown in the top two panels for the 7.9 μm (sensing stratospheric methane) and 12.3 μm (sensing stratospheric ethane) observations. The Cassini simulation shows that the effects of the severe equatorial cooling should be clearly seen, provided the temporal coverage of the ground-based record was sufficient. VLT/VISIR was one of the few mid-IR instruments in operation at the time, and provided images (**Supplemental Figure 12**) to compare to the simulation. We compute the Orton-index for the QPO from the Cassini results – namely the difference between 15.5° and 3.5° (dotted lines in top panels) for the northern (red curves in (c) and (d)) and southern (blue curves) hemispheres. These Cassini curves are compared to the Orton index extracted from the VLT and Subaru measurements themselves (points with error bars) in the northern hemisphere (red, VISIR only) and southern hemisphere (blue for VISIR; light blue for COMICS). Saturn's ring obscuration means that only southern observations were possible pre-equinox, and only northern observations post-equinox. The dashed sinusoidal wave in panel (d) is a comparison to the expected time series of Orton et al. [3], suggesting that the equatorial contrast has remained anomalous since the 2011 disruption. These charts confirm that changes at 7.9- and 12.3-μm are not in phase with one another, with changes (unrelated to the storm) happening ~3 years earlier in the 12.3-μm filter which typically senses deeper altitudes in the stratosphere.

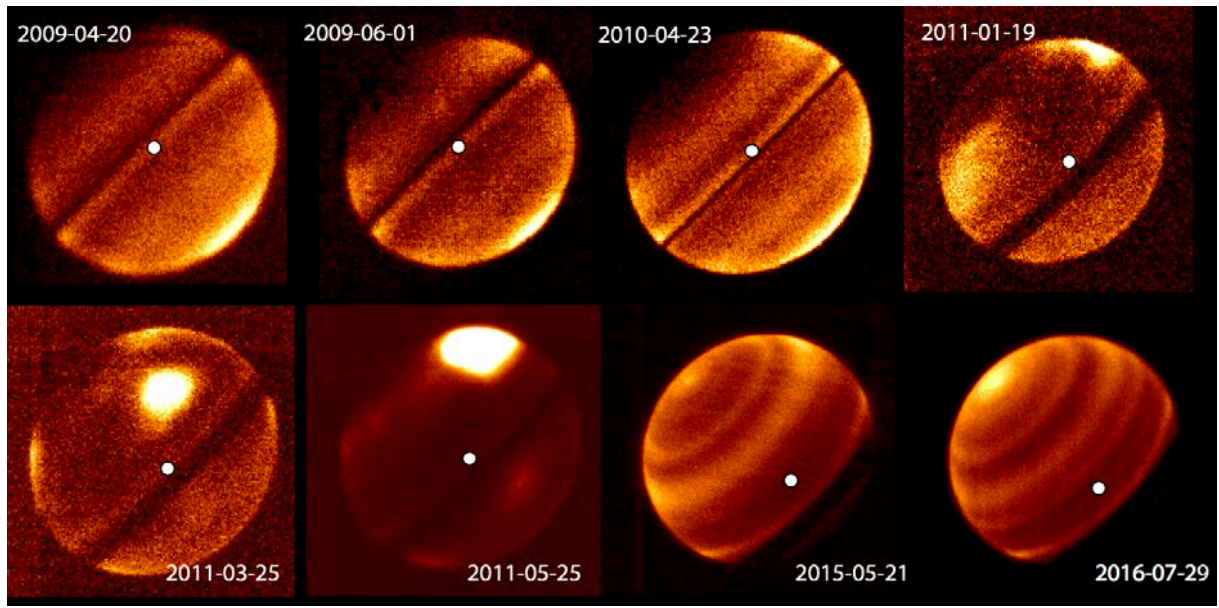

**Supplemental Figure 12:** Examples of VLT/VISIR 7.9-μm imaging record of Saturn from 2009 to 2016. The location of the equator is shown by the white dot. The bright equatorial band is visible in 2009-10 and faintly at the start of 2011, but the growing intensity of the northern stratospheric vortex in March-May 2011 made it difficult to measure contrasts in the equatorial region. VISIR was unavailable from 2012-2014, but annual observations since 2015 (with a much-improved sensitivity and detector plate scale) show the absence of the warm equatorial band as the QPO is in its westward phase. Data were reduced, remapped and calibrated following techniques described by Fletcher et al. [34]. Note the growing brightness of Saturn's north polar stratosphere as the planet approaches summer solstice in May 2017, and the continued presence of a warm homogeneous band at Saturn's northern mid-latitudes as a remnant of the storm. This article considers VISIR data from 2008 (program ID: 381.C-0560), 2009 (383.C-0164), 2010 (084.C-0193) and 2011 (287.C-5032), and then more recently in 2015 (095.C-0142) and 2016 (097.C-0226).

**Supplemental Material (Videos)**

The Cassini CIRS time series displayed in Figs. 1-3 of the main article are better represented as animations. Four animations are provided:

**Video 1 (QPO_temperatures.gif): Evolution of Saturn's equatorial monthly zonal mean temperatures [Kelvin] from 2004 to 2017**. Contours are spaced every 2.5-K, and CIRS has maximum vertical sensitivity in the 0.5-5 mbar and 80-250 mbar range. Temperatures at other altitudes are a combination of interpolation between the troposphere and stratosphere, and smooth relaxation to the *a priori*. Snapshots from this animation are provided in Fig. 2.

https://static-content.springer.com/esm/art%3A10.1038%2Fs41550-017-0271-5/MediaObjects/41550_2017_271_MOESM2_ESM.gif

**Video 2 (QPO_tempanomaly.gif): Evolution of the zonal mean temperature anomaly at Saturn's equator from 2004 to 2017.** A time-averaged temperature profile was subtracted from the measurements in Video 1 at each latitude, revealing the downward propagation of the QPO structure prior to the storm outbreak, and the substantial disruption of the QPO pattern between 2011-2014 at the 0.5-5.0 mbar level. Extratropical counterparts to the QPO pattern are readily seen from ±15-20°, and the heating influence of the storm is evident near 30°N. Contours are spaced every 2 K.

https://static-content.springer.com/esm/art%3A10.1038%2Fs41550-017-0271-5/MediaObjects/41550_2017_271_MOESM3_ESM.gif

**Video 3 (QPO_windshear.gif): Zonal windshear in m/s/km determined from the meridional temperature gradients in Video 1.** Although the modified thermal wind equation (TWE) is used in the main article, the du/dz estimated from the standard TWE is shown here, revealing how the meridional temperature gradients provide vertical shear on the eastward and westward winds. Contours are spaced every 0.5 m/s/km.

https://static-content.springer.com/esm/art%3A10.1038%2Fs41550-017-0271-5/MediaObjects/41550_2017_271_MOESM4_ESM.gif

**Video 4 (QPO_zonalwind.gif): Zonal winds relative to the 5-mbar level estimated from the modified thermal wind equation.** Given the lack of information in the CIRS temperature inversions in the 5-80 mbar range, and the low vertical resolution of the nadir inversions using 15-cm$^{-1}$ resolution spectra, we employ the technique of Flasar et al. (2005), Fouchet et al. (2008), Li et al. (2008) and Guerlet et al. (2011) to estimate the perturbation to the mid-stratospheric winds caused by the measured windshears. Note that these winds are relative to the unknown flow at the 5-mbar level, and do not necessarily imply a switch to retrograde winds.

https://static-content.springer.com/esm/art%3A10.1038%2Fs41550-017-0271-5/MediaObjects/41550_2017_271_MOESM5_ESM.gif